\documentclass{PoS}
\newcommand{\eq}[1]{Eq.~(\ref{#1})}
\title{Tauonic $B$ decays in 2HDMs and the MSSM in the decoupling limit}
\usepackage{amsmath}
\usepackage{amssymb}
\ShortTitle{2HDM III}

\author{\speaker{Andreas Crivellin}
\thanks{This work is supported by the Swiss National Science Foundation (SNF). A.C. thanks the organizing committee for partial support.}\\
        Albert Einstein Center for Fundamental Physics, \\Institute for Theoretical Physics, University of Bern.\\
        E-mail: \email{crivellin@itp.unibe.ch}}

\abstract{In these proceedings we review two aspects of 2HDMs with generic Yukawa structures. The first part considers how recent deviations from the SM expectations in tauonic $B$ decays (observed by BABAR) can be explained in a 2HDM of type III with sizable flavour violation in the up-sector~\cite{Crivellin:2012ye}. The second part discusses the matching of the MSSM on the 2HDM of type III. Here we focus on the recently calculated two-loop SQCD corrections to the Higgs-quark-quark couplings~\cite{Crivellin:2012zz}.}

\FullConference{Prospects for Charged Higgs Discovery at Colliders - Charged 2012,\\
		October 8-11, 2012\\
		Uppsala University, Sweden}

\begin{document}

\section{\label{sec:level1}Introduction}

The SM contains only one scalar isospin doublet, the Higgs doublet. After electroweak symmetry breaking, this gives masses to up quarks, down quarks and charged leptons. The charged component of this doublet becomes the longitudinal component of the $W$ boson and the neutral CP-odd component becomes the longitudinal component of the $Z$ boson. Thus we have only one physical neutral Higgs particle. In a 2HDM \cite{Lee:1973iz} we introduce a second Higgs doublet and obtain four additional physical Higgs particles (in the case of a CP conserving Higgs potential): the neutral CP-even Higgs $H$, a neutral CP-odd Higgs $A$ and the two charged Higgses $H^{\pm}$. The most general Lagrangian for the Yukawa interactions (which corresponds to the 2HDM of type III) in the physical basis with diagonal quark mass matrices is given by
\begin{eqnarray}
\renewcommand{\arraystretch}{2.2}
\begin{array}{l}
\mathcal{L}^{eff}  = \bar u_{f\;L}^{} V_{fj} \left( {\dfrac{{m_{d_i} }}{{v_d }}\delta_{ij}H_d^{2\star}  - \epsilon_{ji}^{d} \left( {H_u^1  + \tan \left( \beta  \right)H_d^{2\star} } \right)} \right) d_{i\;R}  \\ 
\phantom{\mathcal{L}^{eff}  =}  
+ \bar d_{f\;L} V_{j f}^{\star} \left( {\dfrac{{m_{u_j} }}{{v_u }}\delta_{ij}H_u^{1\star}  - \epsilon_{ji}^{u} \left( {H_d^2  + \cot \left( \beta  \right)H_u^{1\star} } \right)} \right) u_{i\;R}  \\ 
\phantom{\mathcal{L}^{eff}  =}  
- \bar d_{f\;L}  \left( {\dfrac{{m_{d_i} }}{{v_d }}\delta_{fi}H_d^{1\star}  + \epsilon_{fi}^{d} \left( {H_u^2  - \tan \left( \beta  \right)H_d^{1\star} } \right)}  \right) d_{i\;R}  \\ 
\phantom{\mathcal{L}^{eff}  =}  - \bar u_{f\;L}^a \left( {\dfrac{{m_{u_i} }}{{v_u }}\delta_{fi}H_u^{2\star}  + \epsilon_{fi}^{u} \left( {H_d^1  - \cot \left( \beta  \right)H_u^{2\star} } \right)} \right)u_{i\;R} \,+\,h.c.  \\ 
 \end{array}
\label{L-Y-FCNC}
\end{eqnarray}
where $\epsilon^q_{ij}$ parametrizes the non-holomorphic corrections which couple up (down) quarks to the down (up) type Higgs doublet\footnote{Here the expression ``non-holomorphic" already implicitly refers to the MSSM where non-holomorphic couplings involving the complex conjugate of a Higgs field are forbidden.}. In the MSSM at tree-level $\epsilon^q_{ij}=0$, which also corresponds to the 2HDM of type II, and flavour changing neutral Higgs couplings are absent. However, at the loop-level, the non-holomorphic couplings $\epsilon^q_{ij}$ are generated~\cite{Hamzaoui:1998nu}.

In these proceedings we consider two different aspects of the 2HDM of type III. In the next section we focus on tauonic $B$ decays in this model and in Sec.~\ref{sec:MSSM-2HDM} we discuss the matching of the MSSM on the 2HDM of type III at NLO in $\alpha_s$.

\section{Tauonic $B$ decays in the 2HDM of type III}

Tauonic $B$-meson decays are an excellent probe of new physics: they test lepton flavor universality satisfied in the Standard Model (SM) and are sensitive to new particles which couple proportionally to the mass of the involved particles (e.g. Higgs bosons) due to the heavy $\tau$ lepton involved. Recently, the BABAR collaboration performed an analysis of the semileptonic $B$ decays $B\to D\tau\nu$ and $B\to D^*\tau\nu$ using the full available data set \cite{BaBar:2012xj}. They find for the ratios
\begin{equation}
{\cal R}(D^{(*)})\,=\,{\cal B}(B\to D^{(*)} \tau \nu)/{\cal B}(B\to D^{(*)} \ell \nu)\,,
\end{equation}
the following results:
\begin{eqnarray}
{\cal R}(D)\,=\,0.440\pm0.058\pm0.042  \,,\\
{\cal R}(D^*)\,=\,0.332\pm0.024\pm0.018\,.
\end{eqnarray}
Here the first error is statistical and the second one is systematic. Comparing these measurements to the SM predictions
\begin{eqnarray}
{\cal R}_{\rm SM}(D)\,=\,0.297\pm0.017 \,, \\
{\cal R}_{\rm SM}(D^*) \,=\,0.252\pm0.003 \,,
\end{eqnarray}
we see that there is a discrepancy of 2.2\,$\sigma$ for $\cal{R}(D)$ and 2.7\,$\sigma$ for $\cal{R}(D^*)$ and combining them gives a $3.4\, \sigma$ deviation from the SM~\cite{BaBar:2012xj}. This evidence for new physics in $B$-meson decays to taus is further supported by the measurement of $B\to \tau\nu$ 
\begin{equation}
{\cal B}[B\to \tau\nu]=(1.15\pm0.23)\times 10^{-4}\,.
\end{equation}
which disagrees with the SM prediction by $1.6\, \sigma$ using $V_{ub}$ from a global fit of the CKM matrix \cite{Charles:2004jd}.

A natural possibility to explain these enhancements compared to the SM prediction is a charged scalar particle which couples proportionally to the masses of the fermions involved in the interaction: a charged Higgs boson. A charged Higgs  affects $B\to \tau\nu$~\cite{Hou:1992sy}, $B\to D\tau\nu$ and $B\to D^*\tau\nu$~\cite{Tanaka:1994ay}. 

In a 2HDM of type II (with MSSM like Higgs potential) the only free additional parameters are $\tan\beta=v_u/v_d$ (the ratio of the two vacuum expectation values) and the charged Higgs mass $m_{H^\pm}$ (the heavy CP even Higgs mass $m_{H^0}$ and the CP odd Higgs mass $m_{A^0}$ can be expressed in terms of the charged Higgs mass and differ only by electroweak corrections). In this setup the charged Higgs contribution to $B\to \tau\nu$ interferes necessarily destructively with the SM \cite{Hou:1992sy}. Thus, an enhancement of $\cal B\left[B\to \tau\nu\right]$ is only possible if the absolute value of the charged Higgs contribution is bigger than two times the SM one\footnote{Another possibility to explain $B\to \tau\nu$ is the introduction of a right-handed $W$-coupling \cite{Crivellin:2009sd}.}. Furthermore, a 2HDM of type II cannot explain $\cal{R}(D)$ and $\cal{R}(D^*)$ simultaneously \cite{BaBar:2012xj}.

\begin{figure}[t]
\centering
\includegraphics[width=0.6\textwidth]{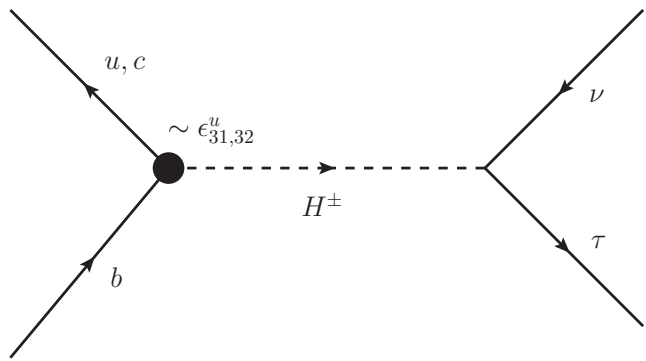}
\caption{Feynman diagram with a charged Higgs contributing to $B\to \tau\nu$ and $B\to D^{(*)}\tau\nu$. The dot represents the flavor-violating interaction containing the 2HDM of type III parameters $\epsilon^u_{31}$ and $\epsilon^u_{32}$, which affect $B\to \tau\nu$ and $B\to D^{(*)}\tau\nu$, respectively.
\label{feynman-diagram}}
\end{figure}

\begin{figure*}[t]
\centering
\includegraphics[width=0.3\textwidth]{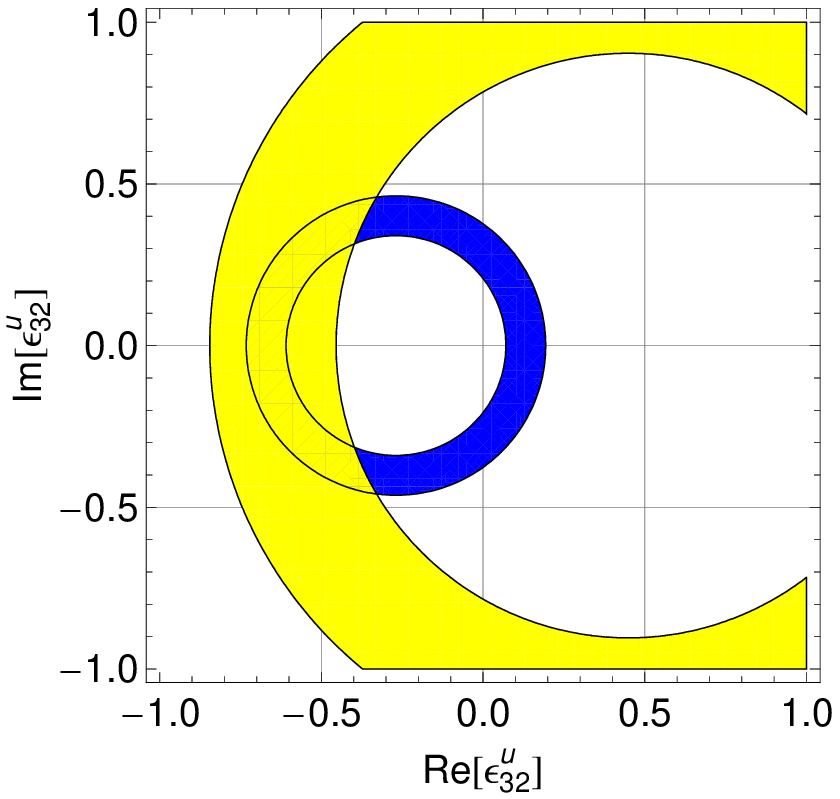}
\includegraphics[width=0.31\textwidth]{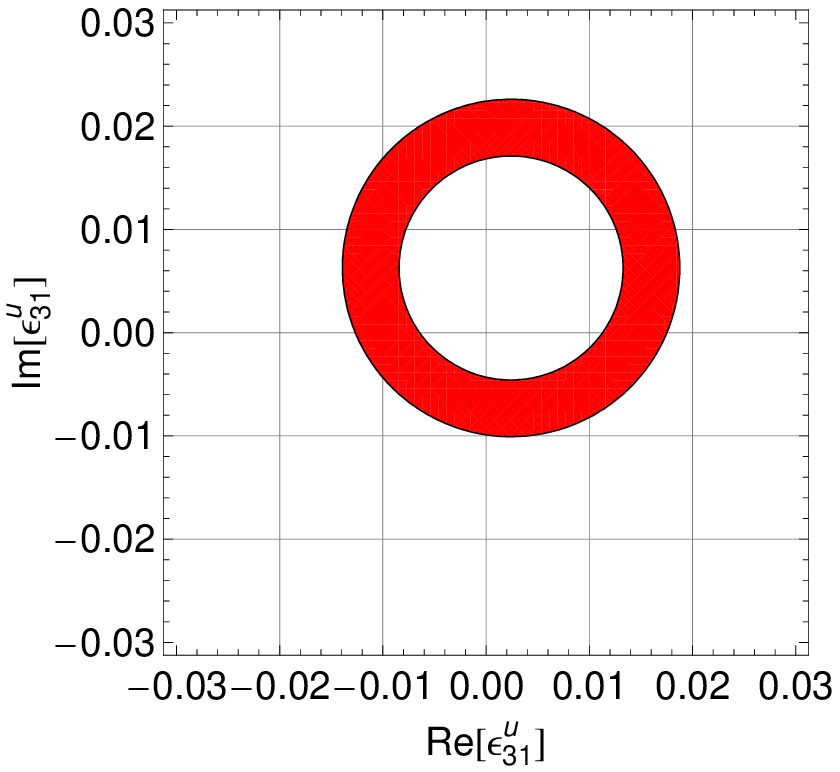}
\includegraphics[width=0.31\textwidth]{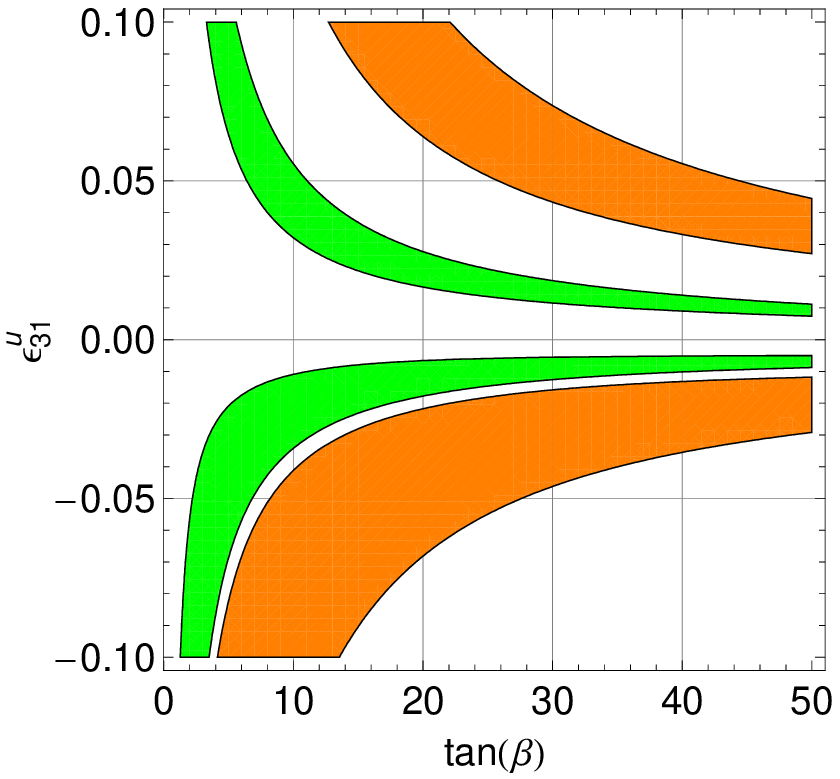}
\caption{Left: Allowed regions in the complex $\epsilon^u_{32}$-plane from $\cal{R}(D)$ (blue) and $\cal{R}(D^*)$ (yellow) for $\tan\beta=50$ and $m_H=500$~GeV. Middle:  Allowed regions in the complex $\epsilon^u_{31}$-plane from $B\to \tau\nu$. Right:  Allowed regions in the $\tan\beta$--$\epsilon^u_{31}$ plane from $B\to \tau\nu$ for real values of $\epsilon^u_{31}$ and $m_H=400$~GeV (green), $m_H=800$~GeV (orange). The scaling of the allowed region for $\epsilon^u_{32}$ with $\tan\beta$ and $m_H$ is the same as for $\epsilon^u_{31}$. $\epsilon^u_{32}$ and $\epsilon^u_{31}$ are given at the matching scale $m_H$. \label{2HDMIII}}
\end{figure*}

In the 2HDM of type III we have much more free parameters ($\epsilon^q_{ij}$) which can affect the tauonic $B$ decays. First, note that all flavor-changing elements $\epsilon^d_{ij}$ are stringently constrained from FCNC processes in the down sector because of tree-level neutral Higgs exchange. Thus, they cannot have any significant impact on the decays we are interested in, and therefore we are left with~$\epsilon^d_{33}$. Concerning the elements $\epsilon^u_{ij}$ we see that only $\epsilon^u_{31}$ ($\epsilon^u_{32}$) significantly effects $B\to \tau\nu$ ($\cal{R}(D)$ and $\cal{R}(D^*)$) without any CKM suppression. Furthermore, since flavor-changing top-to-up (or charm) transitions are not measured with sufficient accuracy, we can only constrain these elements from charged Higgs-induced FCNCs in the down sector. However, since in this case an up (charm) quark always propagates inside the loop, the contribution is suppressed by the small Yukawa couplings of the up-down-Higgs (charm-strange-Higgs) vertex involved in the corresponding diagrams. Thus, the constraints from FCNC processes are weak, and $\epsilon^u_{32,31}$ can be sizable. Of course, the lower bounds on the charged Higgs mass for a 2HDM of type II from $b\to s\gamma$ of 380~GeV \cite{Hermann:2012fc} must still be respected by our model (unless $\epsilon^u_{23}$ generates a destructively interfering contribution), and also the results from direct searches at the LHC for $H^0,A^0\to\tau^+\tau^-$ \cite{Chatrchyan:2012vp} are principle unchanged (if $\epsilon^\ell_{33}$ is not too large). 

Indeed, it turns out that by using $\epsilon^u_{32,31}$ we can explain $\cal{R}(D^*)$ and $\cal{R}(D)$ simultaneously. In Fig.~\ref{2HDMIII} we see the allowed region in the complex $\epsilon^u_{32}$-plane, which gives the correct values for $\cal{R}(D)$ and $\cal{R}(D^*)$ within the $1\, \sigma$ uncertainties for $\tan\beta=50$ and $M_H=500$~GeV. Similarly, $B\to \tau\nu$ can be explained by using $\epsilon^u_{31}$.

\section{Effective Higgs Vertices in the MSSM}
\label{sec:MSSM-2HDM}

In this section we discuss the matching of the MSSM on the 2HDM considering the Yukawa sector but neglecting loop-corrections to the Higgs potential. At tree-level, the MSSM is a 2HDM of type II but at the loop-level, the Peccei Quinn symmetry of the Yukawa sector is broken by terms proportional to the higgsino mass parameter $\mu$ (or non-holomorphic $A^\prime$ terms). 

In the MSSM there is a one-to-one correspondence between Higgs-quark-quark couplings and chirality changing quark self-energies (in the decoupling limit\footnote{The non-decoupling corrections are found to be very small \cite{Crivellin:2010er}.}): The Higgs-quark-quark coupling can be obtained by dividing the expression for the self-energy by the vev of the corresponding Higgs field. 

Let us denote the contribution of the quark self-energy with squarks and gluinos to the operator $\overline{q}_f P_R q_i$ by $C_{f i }^{q\,LR}$. It is important to note that this Wilson coefficient is linear in $\Delta^{q\,LR}$, the off-diagonal element of the squark mass matrix connecting left-handed and right-handed squarks. For down squarks we have
\begin{equation}
	\Delta^{d\,LR}_{ij}=-v_d A^d_{ij}-v_u \mu Y^{d_i}\delta_{ij}\,,
\end{equation}
where the term $v_d A^d_{ij}$ originates from a coupling to $H^d$ while the term $v_u \mu Y^{d_i}$ stems from a coupling to $H^u$ (and similarly for up-squarks). Thus we denote the piece of $\hat C_{f i }^{d\,LR}$ involving the $A$-term by $\hat C^{d\,LR}_{fi\,A}$ and the piece containing $v_u \mu Y^{d_i}$ by $\hat C^{\prime\, d\,LR}_{fi}$. We now define 
\begin{equation}
\renewcommand{\arraystretch}{2}
\begin{array}{l}
   \hat E^d_{fi}\,=\,\dfrac{\hat C^{d\,LR}_{fi\,A}}{v_d}\,,\hspace{1.5cm} 
   \hat E^{\prime d}_{fi}\,=\,\dfrac{\hat C^{\prime\, d\,LR}_{fi}}{v_u}\,,\hspace{1.5cm} 
   \hat E^u_{fi}\,=\,\dfrac{\hat C^{u\,LR}_{fi\,A}}{v_u}\,,\hspace{1.5cm} 
   \hat E^{\prime u}_{fi}\,=\,\dfrac{\hat C^{\prime\, u\,LR}_{fi}}{v_d}\,,
   \end{array}
   \label{E-Sigma}
\end{equation}
where the parameters $\hat E_{fi}^{q}$ ($\hat E_{fi}^{\prime q}$) correspond to (non-)holomorphic Higgs-quark couplings. With these conventions, the couplings  $\epsilon^q_{ij}$ of the 2HDM in \eq{L-Y-FCNC} can be related to MSSM parameters
\begin{eqnarray}
\renewcommand{\arraystretch}{2.0}
\begin{array}{l}
 \epsilon_{fi}^{q}  = \hat E_{fi}^{\prime q}  - \left( 
 {\begin{array}{*{20}c}
   
   0 & 
   {\hat E_{22}^{\prime q} \dfrac{\hat C_{12}^{q\,LR}}{m_{q_2}} } 
   & \hat E_{33}^{\prime q} \left(  \dfrac{\hat C_{13}^{q\,LR}}{m_{q_3}}  - \dfrac{\hat C _{12}^{q\;LR}}{m_{q_2}} \dfrac{\hat C_{23}^{q\,LR}}{m_{q_3}}  \right)  \\
   {\hat E_{22}^{\prime q} \dfrac{\hat C _{21}^{q\;LR}}{m_{q_2}} } 
   & 0 
   & {\hat E_{33}^{\prime q} \dfrac{\hat C _{23}^{q\;LR}}{m_{q_3}}}  \\
   {\hat E_{33}^{\prime q} \left( {\dfrac{\hat C _{31}^{q\,LR}}{m_{q_3}}  - \dfrac{\hat C _{32}^{q\;LR}}{m_{q_3}} \dfrac{\hat C _{21}^{q\,LR}}{m_{q_2}} } \right)} 
   & {\hat E_{33}^{\prime q} \dfrac{\hat C _{32}^{q\,LR}}{m_{q_3}} } 
   & 0  \\
\end{array}} \right)_{fi}  \,. 
 \end{array}
\label{Etilde}
\end{eqnarray}

\begin{figure}
\centering
\includegraphics[width=0.49\textwidth]{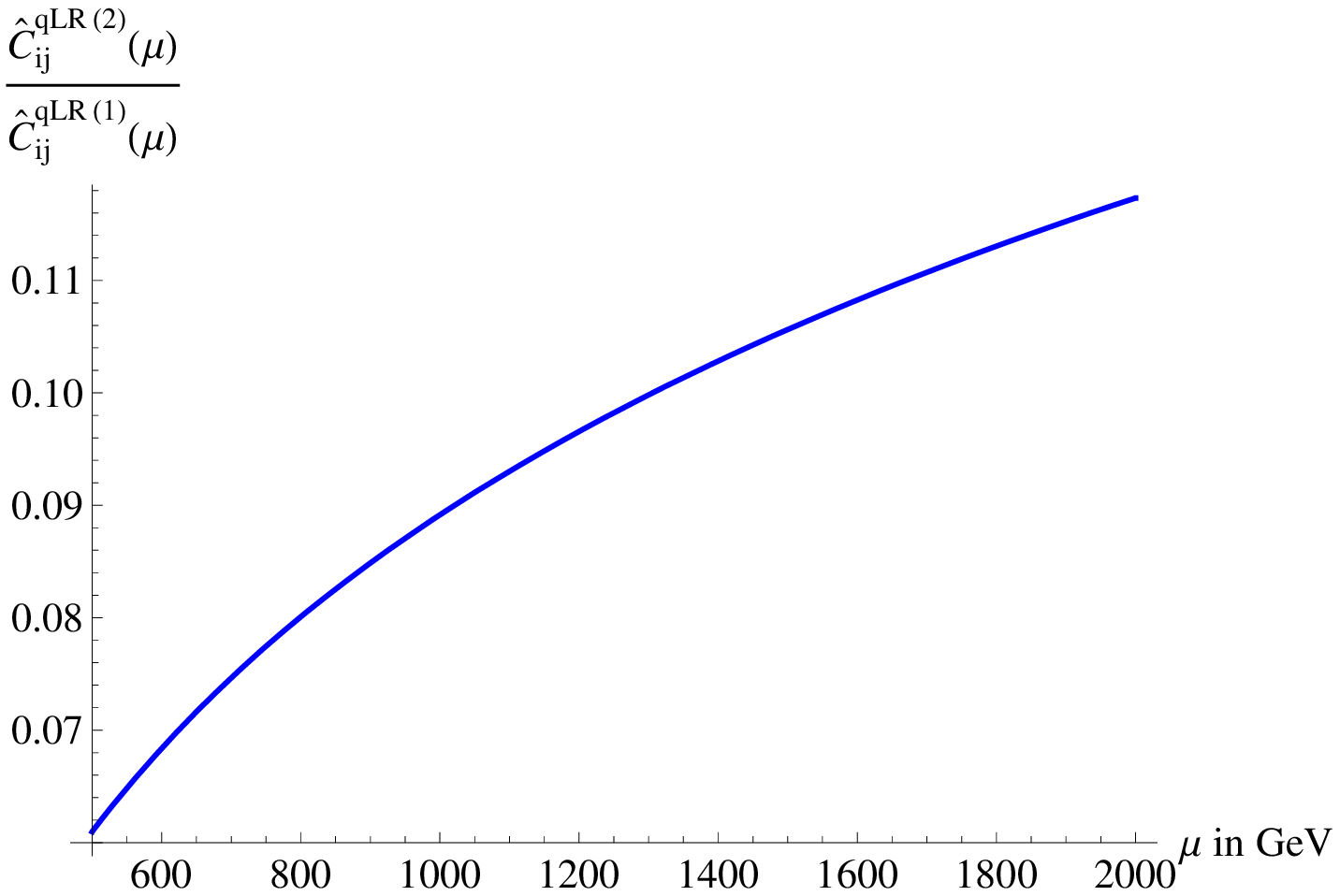}
\includegraphics[width=0.49\textwidth]{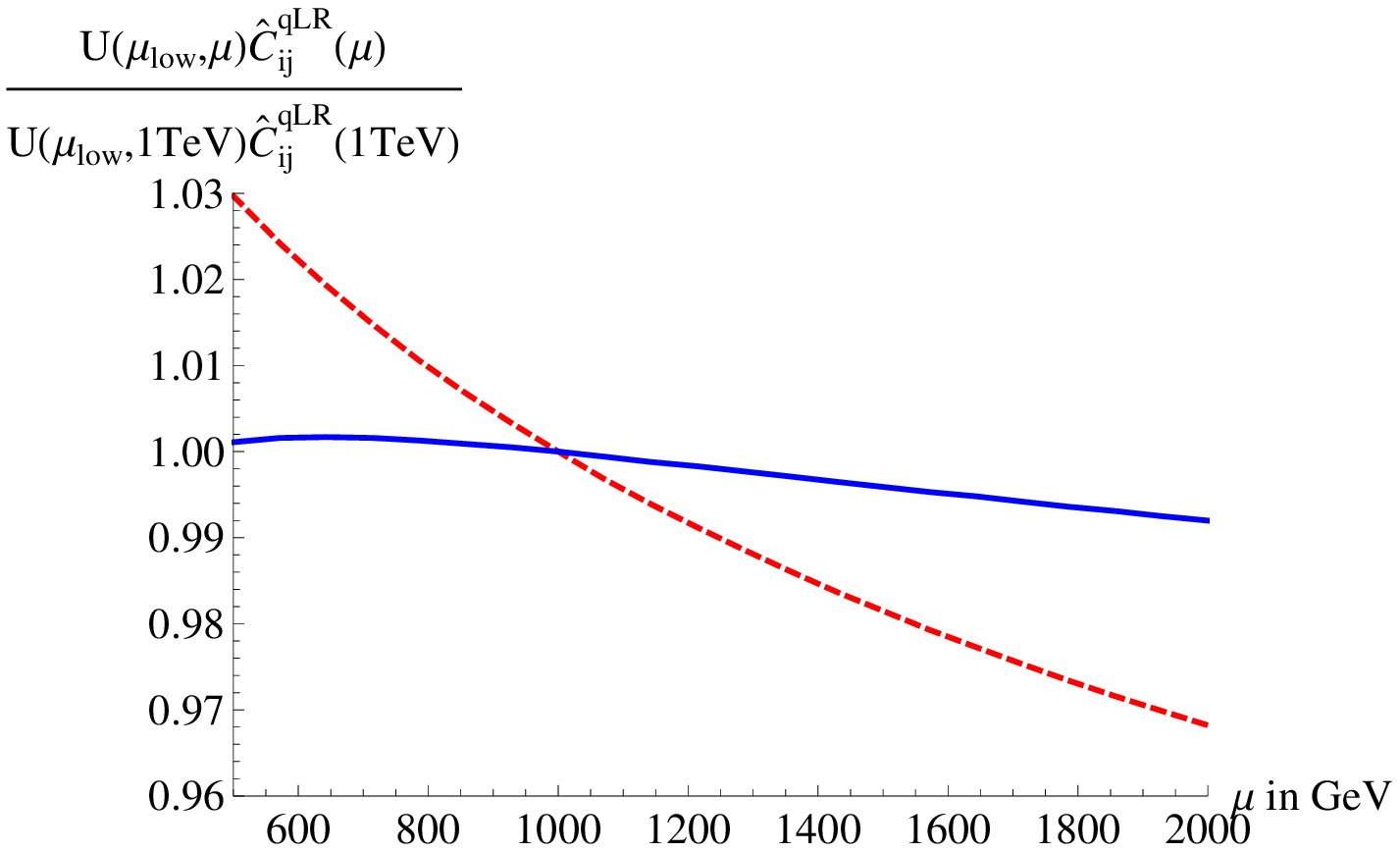}
\caption{Left:Relative importance of the two-loop corrections as a function of the matching scale $\mu$. We see that the two-loop contribution is approximately +9\% of the one-loop contribution for $\mu=M_{\rm SUSY}=1\, {\rm TeV}$.\newline
Right: Dependence on the matching scale $\mu$ of the one-loop and
  two-loop result for $\hat C _{fi}^{q\,LR}(\mu_{\rm low})$, using $M_{\rm SUSY}=1$~TeV and $\mu_{\rm low}=m_W$. Red (dashed): matching done at LO;  blue (darkest):  matching done at NLO matching.  As expected, the
  matching scale dependence is significantly reduced. For the one-loop result, $\hat  C _{fi}^{q\,LR}$ is understood to be $C_{fi}^{q\,LR\,(1)}$ (see text).
  \label{mu-abhaengigkeit}}
\end{figure}

In the matching of the MSSM on the 2HDM one can as a by product also determine the Yukawa couplings of the MSSM superpotential which is important for the study of Yukawa coupling unification in supersymmetric GUTs. Due to this importance of the chirality changing self-energies we calculated them (and thus also $\hat C^{q\,LR}_{ij}$) at the two loop-level in Ref.~\cite{Crivellin:2012zz}. The result is a reduction of the matching scale dependence (see right plot of Fig.~\ref{mu-abhaengigkeit}) while at the same time, the one-loop contributions are enhanced by a relative effect of 9\% (see left plot of Fig.~\ref{mu-abhaengigkeit}). For a numerical analysis also the LO chargino and neutralino contributions should be included by using the results of Ref.~\cite{Crivellin:2011jt}.

Concerning the tauonic $B$-decays discussed in the last section, the size of the quantities $\epsilon^u_{32,31}$ that can be generated via loops in the MSSM is too small to give a sizable effect.

\end{document}